# Bridging coupling bandgaps in nonlinear acoustic metamaterials

Xin Fang*, Jihong Wen†, Dianlong Yu

*Laboratory of Science and Technology on Integrated Logistics Support, National University of Defense Technology, Changsha, Hunan 410073, China.*

Nonlinear acoustic metamaterials (NAMs) open new freedoms in exploiting novel technologies for wave manipulations. Recently, the desired ultra-low and ultra-broad-band wave suppressions were achieved by the chaotic bands in NAMs [Nature Commun. **8**, 1288 (2017)]. This work describes a remote interaction mechanism in NAMs—bridging coupling of nonlinear locally resonant bandgaps. Bridging bandgaps generate chaotic bands and share the negative mass between nonlinear resonators. The bandwidth and the efficiency for the wave reduction in chaotic bands can be manipulated effectively by modulating the frequency distance between the bridging pair. Theoretical analyses on the triatomic model containing two nonlinearly coupled resonances clarify the principle of bridging bandgaps. NAM beams are created to demonstrate this mechanism experimentally by including the bifurcations of periodic solutions. Our study extends the content of NAMs and more nonlinear effects are anticipated based on this mechanism.

## I. INTRODUCTION

Acoustic metamaterials (AMs) provide unconventional and unusual approaches to control subwavelength elastic/acoustic waves [1]. In 2000, the phononic crystal with embedded locally resonant (LR) objects [2] was demonstrated featuring negative effective mass ($m_{\text{eff}}$) at subwavelength scale [3]. Since then, characteristics, underlying physics and applications of AMs have attracted extensive attention [4]. Negative $m_{\text{eff}}$, negative effective bulk modulus [5] ($E_{\text{eff}}$) and double-negative parameter [6-9] have been realized using LR mechanism. LR bandgaps induced by this regime suppress the low-frequency wave propagations [1]. At present, the vast majority of studies have focused on the linear AMs (LAMs) [10]. For LAMs, the linear relationships between the dynamic stress $\sigma_f$ (or concentrated force $F$) and the dynamic strain $\varepsilon$ (or acceleration $a$) are expressed by $\sigma_f = E_{\text{eff}}\varepsilon$ and $F=m_{\text{eff}}a$. In the past decade, diverse functionalities of LAMs are explored, including sound insulation and suppression [11-13], super-absorption [14], super-resolution [15], negative refraction [16], cloaking [17, 18] and phase manipulation using space-coiling metasurfaces [19, 20].

However, LR bandgap is narrow in nature [10] and its generalized bandwidth $\gamma$ (ratio of the bandwidth to its start frequency) depends on the attached mass ratio [1]. Generally, $\gamma$<<1 for single LR bandgap. Moreover, the passbands of finite LAMs consist of dense resonances that amplify waves. These properties limit LAMs' applications in fields like vibration and noise control, where broadband wave suppression is desired. A natural approach to expand the total width of a bandgap is connecting a LR bandgap with Bragg or other LR bandgaps, that is, Bandgap Coupling [21, 22]. The connection is feasible by introducing multiple resonances [23, 24]. The frequency distance between two bandgaps is $\Delta\omega = \omega_{\text{st}}^{(\text{h})} - \omega_{\text{c}}^{(\text{l})}$, herein $\omega_{\text{c}}^{(\text{l})}$ ($\omega_{\text{st}}^{(\text{h})}$) denotes the cutoff frequency of the lower bandgap (the start frequency of the higher bandgap). Bandgap coupling in LAM implies $\Delta\omega \to 0^+$ or $\Delta\omega<0$, so we define it as Adjacent Coupling. Adjacent coupling overcomes the bandwidth of single LR bandgap in a certain extent [21-26], but the total widths are still narrow and resonances in passbands may grow.

* xinfangdr@sina.com
† wenjihong@vip.sina.com

Nonlinearities can boost the development of novel methods for achieving wave manipulations. Extensive studies on nonlinear elastic periodic structures, such as FPU chains [27, 28] and granular crystals [29], have found interesting nonlinear physical phenomena, including solitons [30], amplitude-dependent bandgaps [31] and acoustic diodes [32, 33]. In weakly nonlinear electromagnetic metamaterials, many nonlinear effects such as nonlinear self-action, parametric interactions and frequency conversion were demonstrated [34-37]. However, nonlinear acoustic metamaterial (NAM) is a young topic appearing recently. For the proposed AMs made of side holes, Helmholtz resonators or membranes, weak nonlinearities arise when the intensity of the sound field becomes extremely high [38-40]. These nonlinear acoustic fields in AMs lead to the bandgap shifting and the second harmonic generation [41, 42].

Recently, X. Fang *et al* [43-45] studied the amplitude-dependent dispersion properties, bifurcations, chaos, and the band manipulations in discrete strong NAM models. The wave in the bandgap of NAM features a multi-state behavior, so the LR bandgap becomes nonlinear local resonant (NLR) bandgap [45]. They found the passbands of finite NAMs become chaotic bands, in which a periodic wave becomes chaotic emerging wave with greatly reduced wave transmission [43-45]. Furthermore, by combing the narrow bandgaps and broad chaotic bands under strongly nonlinear state, the desired ultra-low frequency, ultra-broad-band and highly efficient wave suppressions (double-ultra effect) were achieved in the proposed NAM beam ($\gamma$=21) and NAM plate ($\gamma$=39) [46]. Chaotic band opens new avenues in double-ultra control of waves. However, mechanisms giving rise to such a broad chaotic band, especially the interactions between multiple nonlinear resonances in the meta-cells [46], are not comprehensively explained. Moreover, there's great demand to in innovating approaches to manipulate chaotic bands in practice.

In this work, we study the influences of nonlinearly coupled resonators in the periodic cells on the wave propagations in NAMs with at least two LR bandgaps. Different from the adjacent couplings, we find a special remote interaction between NLR bandgaps—bridging coupling—that overcomes the limitation of the frequency distance $\Delta\omega$ and enables effective manipulation of the chaotic band. Bridging couplings of NLR bandgaps give rise to chaotic band. The bandwidth and the efficiency for the wave reduction can be adjusted by the distance. Theoretical analyses on the triatomic chain clarify the principle of bridging bandgaps. One-dimensional NAM beams are fabricated to demonstrate this mechanism by including the bifurcations of periodic solutions.

# II. PRINCIPLE OF THE BRIDGING COUPLING OF NLR BANDGAPS

Triatomic chain is the fundamental model featuring the nonlinear couplings of two resonant bandgaps. As shown in FIG. 1, the triatomic cell consists of a linear primary oscillator $M$-$k_0$, and two local resonators $m_1$-$k_1$ and $m_2$-$k_2$. In a cell, the displacements of $M$, $m_1$ and $m_2$ are $x$, $y$ and $z$, respectively, and $u$=$z$-$y$. The resonators, $m_1$ and $m_2$, are coupled through a linearly viscous damping $c\dot{u}$ and an intrinsic nonlinear spring $k_n u^3$.

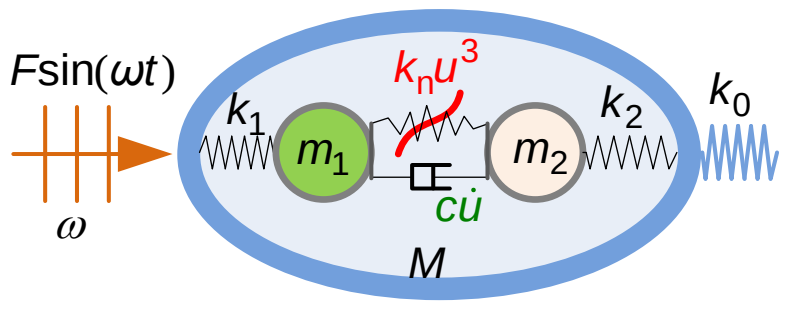

FIG. 1. The triatomic cell. $M$, $m_1$ and $m_2$ are masses of the three oscillators; $k_0$, $k_1$, $k_2$ are their linear stiffnesses.

The motion equations of the $n$th cell in the triatomic chain are

$$\begin{cases} M\ddot{x}_n = k_1(y_n - x_n) + k_2(u_n + y_n - x_n) + F(t) \\ m_1\ddot{y}_n = -k_1(y_n - x) + c\dot{u}_n + k_n u_n^2 \\ m_2(\ddot{u}_n + \ddot{y}_n) = -k_2(u_n + y_n - x_n) - c\dot{u}_n - k_n u_n^2 \end{cases} \quad (1)$$

when calculating the responses of the chain, $F(t)$ $=k_0(x_{n\text{-}1}+x_{n+1}\text{-}2x_n)$. The effective mass $m_{\text{eff}}$ describes interesting properties of acoustic metamaterials. By specifying $F(t)$=$F$·sin$\omega t$, $x$=$X$·sin$\omega t$, $y$=$Y$·sin$\omega t$, $u$=$U$·sin$\omega t$ in Eq. (1), neglecting the damping effect and adopting the first-order harmonic balance method, $m_{\text{eff}}$ of the nonlinear triatomic cell is given by

$$m_{\text{eff}} = \frac{\langle F \rangle}{\langle \ddot{x} \rangle} = \frac{F}{-\omega^2 X} = \alpha \cdot m_{\text{eff}}^{(\text{L})} \quad (2)$$

where $m_{\text{eff}}^{(\text{L})}$ denotes the effective mass of the linear triatomic cell, and $\alpha$ represents the effect of the nonlinearity.

$$m_{\text{eff}}^{(\text{L})} = M + \frac{k_1}{\omega_1^2 - \omega^2} + \frac{k_2}{\omega_2^2 - \omega^2} \tag{3}$$

$$\alpha = \left[ 1 + \frac{3k_{\text{n}} U^3 \omega^2 (\omega_2^2 - \omega_1^2)}{4F(\omega_1^2 - \omega^2)(\omega_2^2 - \omega^2)} \right]^{-1} \tag{4}$$

Here $\omega_i = \sqrt{k_i / m_i}, i = 1,2$ are resonant frequencies and it is specified as $\omega_1 < \omega_2$,. At $\omega_1$ and $\omega_2$, we have

$$m_{\text{eff}}(\omega_1) = \frac{4Fk_1}{3k_{\text{n}} U_{(1)}^3 \omega_1^2}, \quad m_{\text{eff}}(\omega_2) = \frac{4Fk_2}{3k_{\text{n}} U_{(2)}^3 \omega_2^2} \tag{5}$$

When considering the viscous dissipation, the linear effective mass is expressed as

$$m_{\text{eff,d}}^{(\text{L})} = M + \frac{k_1(\omega_2^2 - \omega^2) + k_2(\omega_1^2 - \omega^2) - ic\left(\frac{k_1}{m_2} + \frac{k_2}{m_1} + \omega_1^2 - \omega_2^2\right)}{(\omega_1^2 - \omega^2)(\omega_2^2 - \omega^2) - ic\left(\frac{\omega_2^2 - \omega^2}{m_1} + \frac{\omega_1^2 - \omega^2}{m_2}\right)} \tag{6}$$

Therefore $\text{Re}[m_{\text{eff,d}}^{(\text{L})}] = m_{\text{eff}}^{(\text{L})}$.

The motion equation of the primary discrete monoatomic chain is $M\ddot{x}_n = k_0(x_{n-1} + x_{n+1} - 2x_n)$. This equation can be rewritten as a difference formula, which can model this chain as a continuum medium whose wave equation is given by

$$M \frac{\partial^2 u}{\partial t^2} = k_0 \frac{\partial^2 u}{\partial s^2} \tag{7}$$

Here $s$ denotes the propagation direction. Under the long-wave approximation, replacing $M$ with $m_{\text{eff,d}}^{(\text{L})}$ leads to the wave equation of the effective LAM. Then, we obtain the effective wave vector $\kappa_{\text{eff}}$ of this LAM,

$$\kappa_{\text{eff}} = \omega \sqrt{m_{\text{eff,d}}^{(\text{L})} / k_0} \tag{8}$$

The imaginary part of $\kappa_{\text{eff}}$, $\text{Im}(\kappa_{\text{eff}})$, approximately shows the dissipative characteristic of the effective LAM. A larger $\text{Im}(\kappa_{\text{eff}})$ leads to a stronger attenuation of wave.

To study the properties of effective mass $m_{\text{eff}}/M$ and the responses of the finite chain, we specify parameters $M$=1, $m_1$=0.3, $m_2$=0.3, $\omega_0=(k_0/M)^{1/2}$=6, $\omega_1$=2. The distance between the two bandgaps is controlled by $\omega_2$. $\omega_2$=4 and $\omega_2$=8 are taken as examples. As illustrated in FIG. 2a,b, two narrow bands of negative $m_{\text{eff}}/M$ are generated near the resonant frequencies $\omega_1$ and $\omega_2$ in the linear metamaterials. This regime brings two LR bandgaps, LR1 and LR2.

Equation (2) implies that $m_{\text{eff}}$ of a nonlinear metamaterial is manipulated by the amplitude $U^3/F$, where the relative displacement $U$ depends on the input force $F$. Therefore, $m_{\text{eff}}/M$ near a resonant frequency depends on the motion of the other resonator. There are still two bands for $m_{\text{eff}}/M$<0 near $\omega_1$ and $\omega_2$ in the nonlinear triatomic model. However, the nonlinearity shifts the first negative $m_{\text{eff}}/M$ band up to higher frequencies, and a stronger nonlinearity brings a larger deviation to $\omega_1$. As well established, this is also the process that LR1 becomes NLR bandgap, NLR1, in this NAM model [45]. LR2 will also become NLR2 when significant nonlinearity appears. A saddle-node bifurcation frequency $\omega_{\text{J1}}$ appears in the interval $\omega_1$-$\omega_2$. This bifurcation gives birth to three branches of $m_{\text{eff}}/M$ in $\omega_{\text{J1}}$-$\omega_2$. Let's label the branch corresponding to the minimum $m_{\text{eff}}$ as branch-1 and the largest as branch-3. Branch-2 connects them. As shown in FIG. 2a, Branch-3 starts at $\omega_{\text{J1}}$ and it is coincident with the linear solution; branch-1 originates from $m_{\text{eff}}\to-\infty$ in NLR1 and monotonously approaches to 0 near NLR2, which implies $m_{\text{eff}}$<0 on the whole branch-1. This branch expands the first negative mass region and connects the two negative mass regions, so that the nonlinear coupling between the two resonators $m_1$ and $m_2$ shares the negative $m_{\text{eff}}$ between NLR1 and NLR2 through the energy transmission. Therefore, NLR1 and NLR2 are coupled. This interaction appears in the case $\Delta\omega > 0$, and it is independent of $\Delta\omega$ in theory (FIG. 2b), which indicates this coupling works in the situation with remote frequency distance. Its properties are different with the adjacent couplings in essence. In fact, there are similar bifurcation properties in the band $\omega$>$\omega_2$. However, its branch-1 and branch-2 are not departed and much stronger nonlinearity is required to shift NLR2.

A finite chain made of 20 cells is adopted to describe the response properties. The left boundary of the chain is $x_0=A_0\sin\omega t$, and the right is free. $x_0$ also acts as the driving displacement. To make the comparison be more rigorous, a dissipation $c$=0.01 is considered. Numerical integration method is employed to directly solve the responses in the time interval 0-300s. $A_{\text{av}}(\omega)$ denotes the average displacement in 150-300s (steady response) of the last oscillator $M$. By specifying $A_0$=0.005, the nonlinear strength is modulated by the coefficient $k_n$.

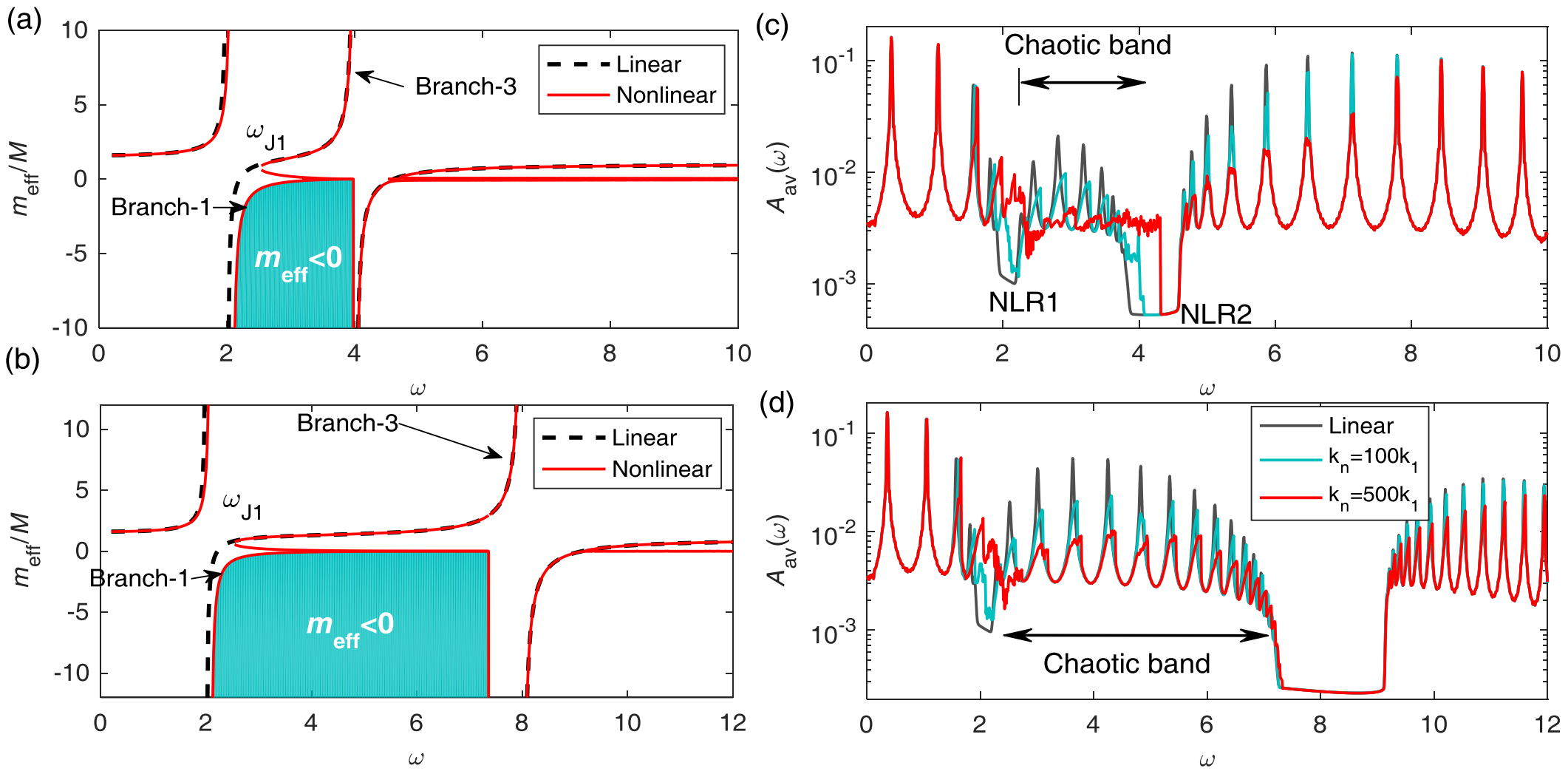

FIG. 2 (a,b) The effective mass $m_{eff}/M$ and (c, d) the responses of the finite chain composing of 20 cells. (a, c) $\omega_2$=4; (b, d) $\omega_2$=8. $F$=0.02 and $k_n$=500$k_1$ when calculating $m_{eff}/M$. $A_0$=0.005 and $c$=0.01 when solving $A_{av}(\omega)$.

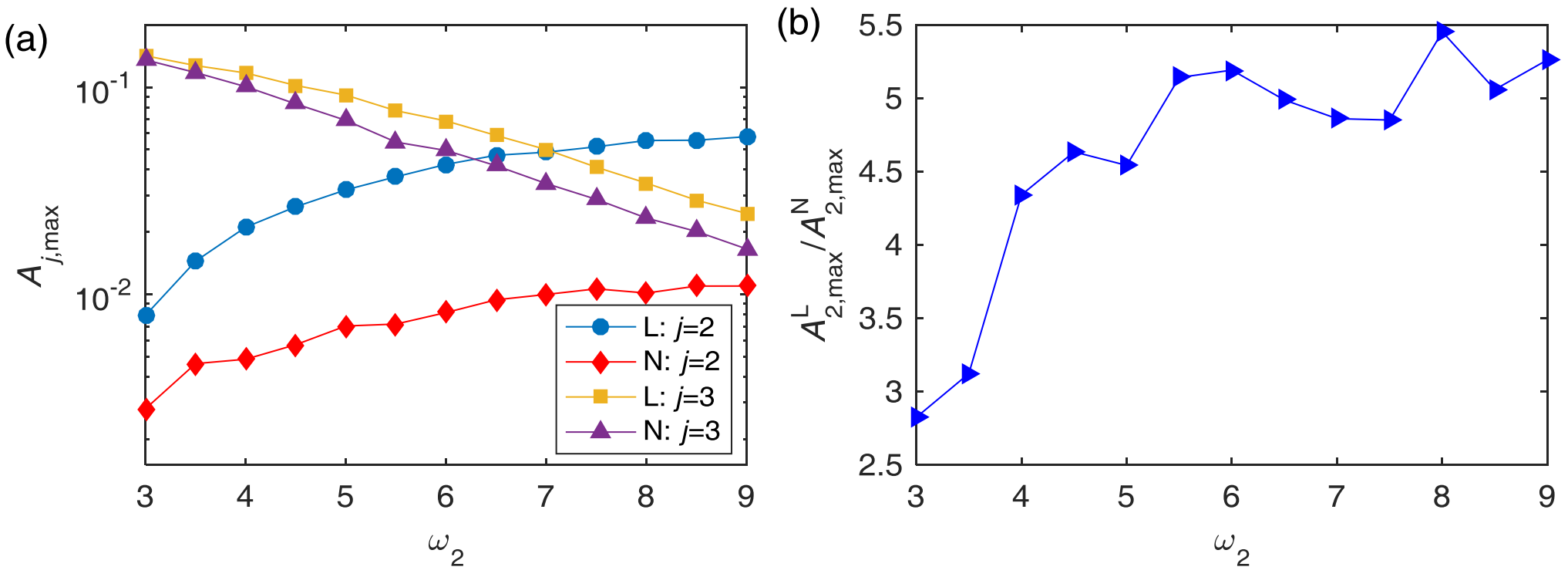

FIG. 3 Response amplitudes of the finite triatomic chain. $A_0$=0.005, $k_n$=500$k_1$. (a) $A_{2,max}$ and $A_{3,max}$ in linear ('L') and nonlinear ('N') cases. (b) The ratio $A^{L}_{2,max}/A^{N}_{2,max}$.

As shown in FIG. 2c,d, the passbands of linear acoustic metamaterial consist of dense resonances, though their amplitudes are limited by the damping. In the case $\omega_2$=4, the maximum amplitude in the second passband is smaller than that in the first and the third passbands. For the given parameters, $k_n$=100$k_1$ and $k_n$=500$k_1$ simulate the weak and moderate nonlinearities, respectively. In contrast with the linear case, the weak nonlinearity can reduce $A_{av}$ at the resonances in the second passband and the lower part of the third passband. Enhancing the nonlinear strength leads to larger amplitude reductions and a broader reduction band. Moreover, $k_n$=500$k_1$ makes NLR1 and the lower part of NLR2 close when $\omega_2$=4. Waves inside the NLR bandgaps are relevant to the bifurcations of fundamental waves and its high-order harmonics. The frequency range in which the resonances are suppressed is the chaotic band. If other parameters are constant, increasing $\omega_2$ to $\omega_2$=8 will broaden the second passbands and LR2, certainly. The interesting phenomenon is that the chaotic band is also expanded under the same nonlinear strength.

Let's define $A_{2,max}$ as the maximum $A_{av}(\omega)$ in the second passband $\omega \in (\omega_c^{(1)}, \omega_{st}^{(2)})$, where $\omega_c^{(1)}$ is the cutoff frequency of NLR1 and $\omega_{st}^{(2)}$ denotes the start frequency of NLR2, that is, $A_{2,max} = \max A_{av}(\omega)$ for $\omega_c^{(1)} < \omega < \omega_{st}^{(2)}$. Similarly, $A_{3,max} = \max A_{av}(\omega)$ for $\omega > \omega_c^{(2)}$. Therefore, $A^{L}_{2,max}/A^{N}_{2,max}$ indicates the influences of the nonlinearity on the responses in the second passband, where the superscripts L and N correspond to the linear and nonlinear cases, respectively.

As shown in FIG. 3, when modulating $\omega_2$ in $3<\omega_2<9$ under the moderate nonlinearity $k_n=500k_1$, we have $A^{L}_{2,\max} \propto \Delta\omega$ and $A^{L}_{3,\max} \propto 1/\Delta\omega$, here $\Delta\omega = \omega_2 - \omega_1$. This law reveals that the dissipative effect of LAMs can also be modulated by the frequency distance of the LR bandgaps. Im($\kappa_{\text{eff}}$) evaluated from Equ. (8) demonstrates this feature. As indicated by FIG. 4, Im($\kappa_{\text{eff}}$)>>$c$=0.01 inside bandgaps; as $\Delta\omega$ increases, Im($\kappa_{\text{eff}}$) in the second passband decreases but Im($\kappa_{\text{eff}}$) in the third passband increases. The dissipative effects of linear resonances follow the same regular with Im($\kappa_{\text{eff}}$), which leads to the behaviors of $A^{L}_{2,\max}$ and $A^{L}_{3,\max}$ characterized in FIG. 3. Moreover, the width of LR2 also increases against $\Delta\omega$.

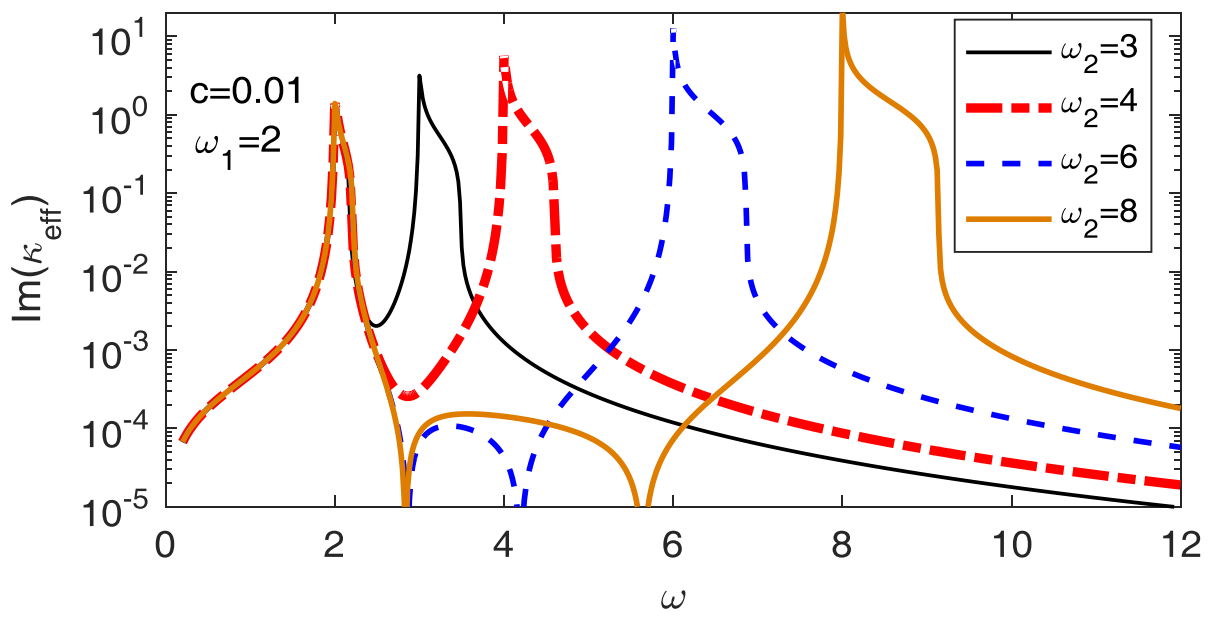

FIG. 4 The imaginary part of the wave vector Im($\kappa_{\text{eff}}$) of the linear triatomic model under damping case. $\omega_1$=2, $c$=0.01.

For the NAM model, there are also $A^{N}_{2,\max} \propto \Delta\omega$ and $A^{N}_{3,\max} \propto 1/\Delta\omega$. However, attributed to the chaotic band effect, $A^{N}_{j,\max} < A^{L}_{j,\max}, j = 2,3$ and $A^{N}_{2,\max}$ increases much slower than $A^{L}_{2,\max}$ with $\Delta\omega$. Accordingly, when $\omega_2$ increases from 3 to 6, $A^{L}_{2,\max}/A^{N}_{2,\max}$ increases from 2.85 to 5.19. Further increasing $\omega_2$ does not alter $A^{L}_{2,\max}/A^{N}_{2,\max}$ obviously. These principles indicate that increasing the distance between two NLR bandgaps in a certain extent can broaden the chaotic band and enhance its efficiency to suppress the resonances. Further increasing the nonlinear strength will bring higher suppressing efficiencies [43-46]. Combining the features in the second passband and the third passband with the width of NLR2, the total bandwidth for wave attenuation and suppression is significantly expanded by increasing $\Delta\omega$. It is noteworthy that these chaotic band's effects manipulated by the distance of NLR bandgaps depend little on the damping, though the damping can also attenuate the wave. This type of interaction between bandgaps arises from the nonlinear effect in nature.

When the finite NAM has two or more NLR bandgaps, analyses above clearly demonstrate that these bandgaps exhibit remote ($\Delta\omega$>>0) couplings by generating the chaotic band together to suppress the resonances significantly; and increasing $\Delta\omega$ in a certain extent under the finite nonlinear strength can increase the suppression efficiency and bandwidth. In this coupling, the bandgaps behave as bridge piers and the chaotic band likes the bridge floors supported by the piers, so we refer to it as Bridging Coupling. Two coupled NLR bandgaps are a Bridging Pair. Bridging coupling is different from the adjacent couplings in essence. Because the nonlinear strength, particular the coefficient $k_n$, realized in a real structure is generally limited by the scale, the amplitude and manufacturing technologies, there is paramount importance using bridging couplings to manipulate the chaotic band effect in practice.

# III. VERIFICATION OF THE BRIDGING COUPLING OF NLR BANDGAPS

## A. Metamaterial design

To verify the bridging coupling of NLR bandgaps, we fabricate NAM beams made of periodic cells shown in FIG. 5a. The primary structure is a uniform linear beam and the lattice constant $a$=80 mm. Two oscillators are fixed on the primary beam through a stiff strut. Oscillator-1 consists of a cantilever thin beam and a tip mass $m_1$. Oscillator-2 consists of a cantilever tube (or a thick beam) and a connector (acting as mass $m_2$) fixed at the end of the tube (or the beam). The first resonant frequency of oscillator-2 is designed higher than that of oscillator-1. There is a square hole in $m_1$. A pair of small pyramid rubbers is attached on the connector and both of them are put in the square hole. The rubbers are undeformed at rest. In theory, the clearance between the peak of the rubber and the wall of the hole is zero. The force $F$ generated by a compressed pyramid rubber varies nonlinearly against its compression deformation $u$. We measure the relationship $F$-$u$ with a high-precision instrument DMA Q800 under the quasi-static compression experiments at about 30 degrees Celsius. As depicted in FIG. 6, the relationship $F$-$u$ induced by a pair of pyramid rubbers can be fit with a cubic nonlinear function $F(u)$ accurately,

$$F(u) = k_0 u + k_n u^3 \tag{9}$$

We realize $k_n=\beta\times10^{10}$ N/m$^3$, $\beta$=1~4, in experiments. In theoretical analyses, we specify $k_n=1\times10^{10}$ N/m$^3$. This $k_n$ enable us achieving moderate nonlinearities with finite driving amplitudes. The linear stiffness $k_0$ is neglectable comparing with oscillators' stiffness coefficients.

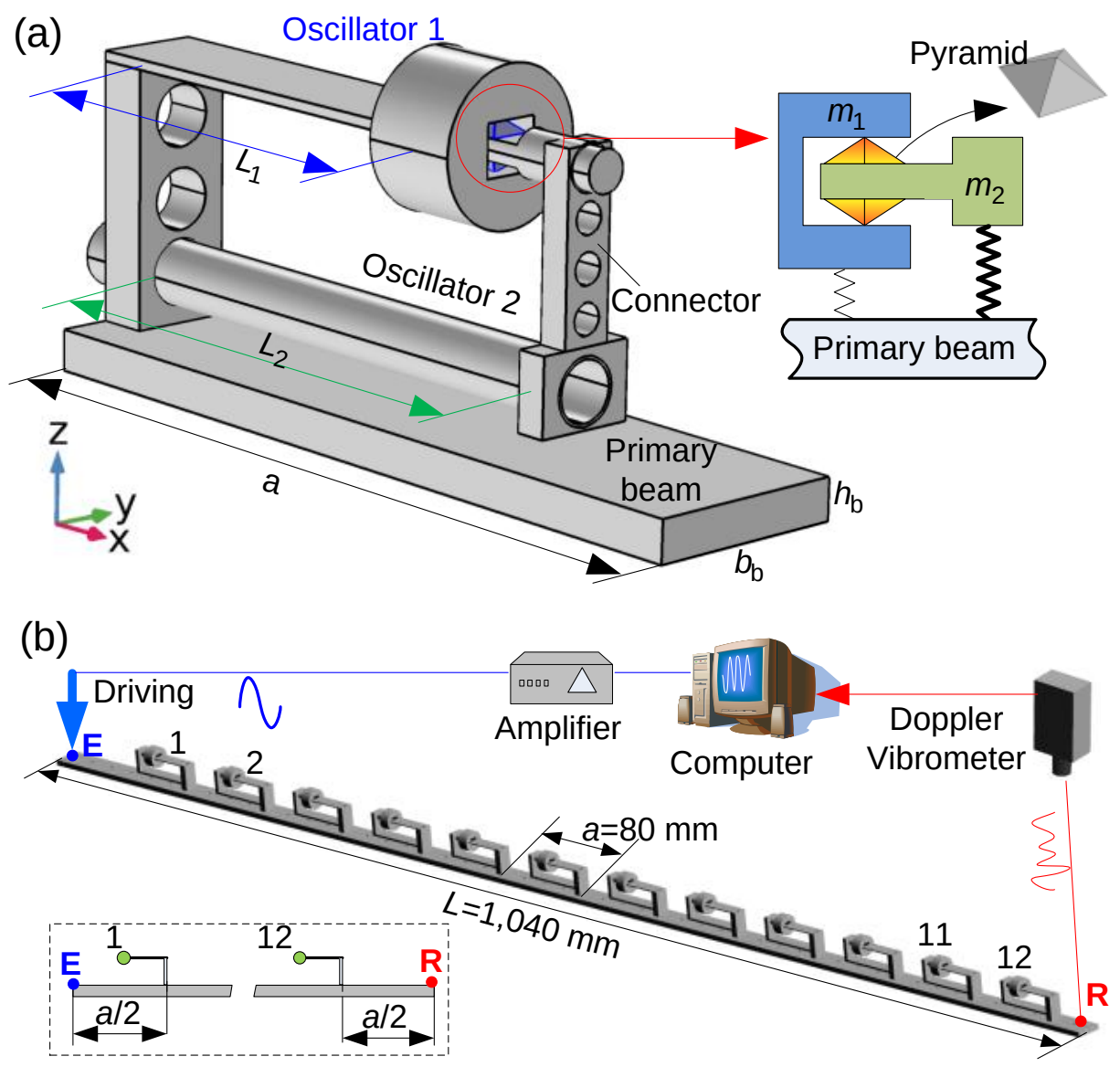

FIG. 5 Structure of the NAM beam. (a) Structure of a cell; the right iconography is the supposed equivalent model. (b) Structure of the NAM beam consists of 12 cells and its measuring method. Points, E and R, are driving and measuring points, respectively. Their positions are labeled in the box.

The whole metamaterial beam consists of 12 cells, as shown in FIG. 5b. The total length of the beam is 1040 mm. In experiments, different levels of broadband white noise signals drive the beam at point E. A laser Doppler vibrometer measures the responses at the right end point R. Driving levels are modulated by the voltage of the amplifier. We fabricate NAM1 and NAM2 to study the influence of frequency distance on the wave transmission of NAMs. The first eigen-frequency of oscillator-2 in NAM2 is higher. NAM1 is shown in FIG. 5b. NAM2 consists of periodic cells shown in FIG. 5a. Their primary beams are same, and its width, thickness, density and modulus of the beam are $b_b$=20 mm, $h_b$=4 mm, $\rho$=2700 kg/m$^3$ and $E$=70 GPa, respectively. The concentrated mass and inertia at the fixed point in a cell is $m_0$=3 g and $J_0$=0, respectively. Parameters of oscillator-1&2 are listed in Table 1. Their definitions are: $\rho_i$—density, $E_i$—modulus, $A_i$—section area, $I_i$—inertia moment, $L_i$—length, $m_{ti}$—tip mass, $J_{ti}$—tip inertia; herein the subscript $i$=1, 2 denotes parameters relevant to oscillator-1 or oscillator-2. In FIG. 5a, inverting the connecter on oscillator-2 decouples (D.C.) oscillator-1 and oscillator-2. On this occasion, they become typical linear metamaterials, LAM1 and LAM2, with multiple local resonances.

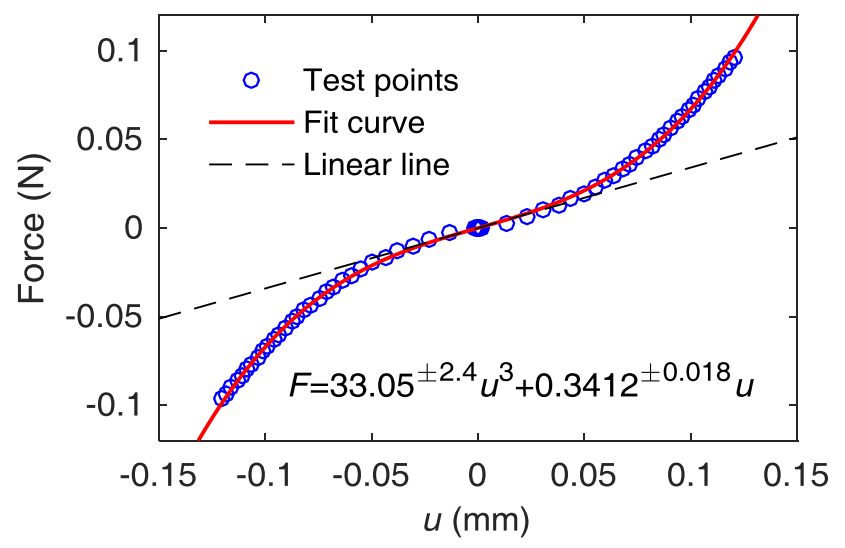

FIG. 6 Nonlinear relations between the compression force $F$ and the deformation $u$ of a pair of pyramid rubbers.

Table 1 Parameters of oscillator-1&2 in NAM1 and NAM2. The subscript $i$=1, 2

| | Oscillator-1 | | Oscillator-2 | |
|---|---|---|---|---|
| Symbol | NAM1 | NAM2 | NAM1 | NAM2 |
| $\rho_i$ (kg/m$^3$) | 2700 | 2700 | 2700 | 7800 |
| $E_i$ (GPa) | 70 | 70 | 70 | 200 |
| $A_i$ (m$^2$) | 8e-6 | 1e-5 | 1.6e-5 | 5.37e-6 |
| $I_i$ (m$^4$) | 4.27e-13 | 8.33e-13 | 3.41e-12 | 1.75e-11 |
| $L_i$ (mm) | 32 | 44.3 | 43.5 | 59.5 |
| $m_{ti}$ (g) | 11.45 | 11.45 | 2.2256 | 4.5 |
| $J_{ti}$ (kg m$^2$) | 2.56e-7 | 2.05e-7 | 1.8e-7 | 5.25e-7 |

## B. Dispersion and transmission properties

By considering the first and the second flexural modes of oscillator-1, and the first flexural mode of oscillator-2, this manuscript establishes the physical models of oscillator-1&2 with mode superposition method (see Appendix). And then, the finite element (FE) model of a periodic cell is built based on the Bloch theorem [46]. Harmonic balance method is adopted to solve the dispersion curves of this model [46]. The dispersion curves of linear and nonlinear metamaterial beams under different amplitudes are illustrated in FIG. 7. LAM1 and LAM2 have four bandgaps below 1400 Hz. Their ranges are listed in Table 2. LR1 and LR3 are induced by the first and the second flexural modes of oscillator-1. LR2 is generated by the first flexural mode of oscillator-2. Both LAM1 and LAM2 have Bragg bandgaps near 1200 Hz. The generalized bandwidth of a band is $\gamma = (f_{cut}-f_{st})/f_{st}$, where $f_{st}$ ($f_{cut}$) denotes the start (cutoff) frequency. LAM2-LR2's $\gamma$ is largest ($\gamma$=0.295), however, $\gamma$<<1. Therefore, those bandgaps are narrow. The main difference between LAM1 and LAM2 lies in the position of LR2. The distance between LR1 and LR2 in LAM2 is farther.

Table 2 Ranges of Bandgaps in LAM1 and LAM2

| | LR1 (Hz) | LR2 (Hz) | LR3 (Hz) | Bragg (Hz) |
|---|---|---|---|---|
| LAM1 | 75.2-90.8 | 275.8-299.3 | 851.9-896.1 | 1151-1286 |
| LAM2 | 64.5-77.9 | 405.2-524.6 | 965.9-1027 | 1072-1270 |

The transfer function is $H(\omega)$=20log$_{10}$[$X_R(\omega)/X_E(\omega)$] dB, where $X_i(\omega)$ is the frequency spectrum at point $i$. $H(\omega)$ of NAM1 and NAM2 under different excitation levels are shown in FIG. 8. The bandgaps, LR1, LR2 and LR3, presented by $H(\omega)$ of the linear decoupled cases agree with the dispersion curves well. The dispersion solutions of NAMs depend on the wave amplitude $A_0$ [43-46]. To provide amplitudes $A_0$ that are consistent with the experiments, the average excitation displacements in 50-100 Hz (near LR1) are adopted as references of $A_0$. $A_0$ also indicates different driving levels. Therefore, we use $A_0$=4 μm and 10 μm for NAM1; and $A_0$=10 μm and 20 μm for NAM2. The smaller $A_0$ approximates the experimental input wave. Besides the five dispersion curves identical with the relevant LAMs, NAMs have multiple dispersion solutions that will bifurcate (see FIG. 7). Most of these multiple solutions have large imaginary parts leading to rapid attenuations of waves. For bandgaps, the nonlinearity firstly shifts the lower boundary of LR1 upwards; further enhancing the nonlinear strength can also shift the lower boundary of LR2. However, their upper boundaries keep almost constant, so as to LR1 closes firstly. Therefore, LR bandgaps become NLR gaps in NAMs. These regimes agree with the triatomic model well. Stronger nonlinearity is required to shift LR2 of NAM2 attributing to its higher frequency. It is similar for LR3.

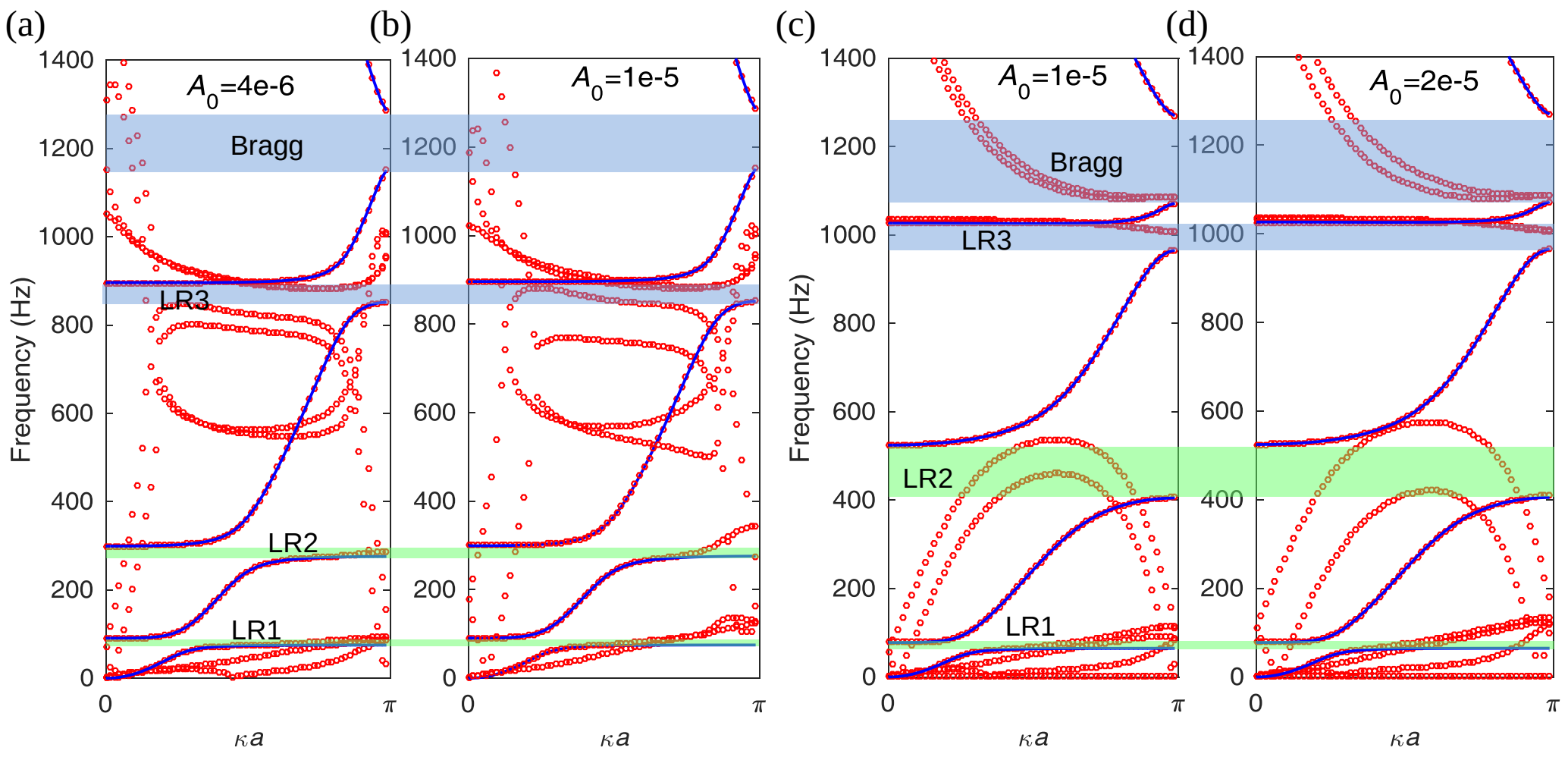

FIG. 7 Dispersion curves of linear and nonlinear metamaterials under different amplitudes. (a, b) NAM1; (c, d) NAM2. Different wave amplitudes $A_0$ are labeled on the corresponding panels. The solid blue curves are linear dispersion results, and all the red points are nonlinear dispersion results. The wave vector $0 < \kappa < \pi/a$. The shaded bands are linear bandgaps.

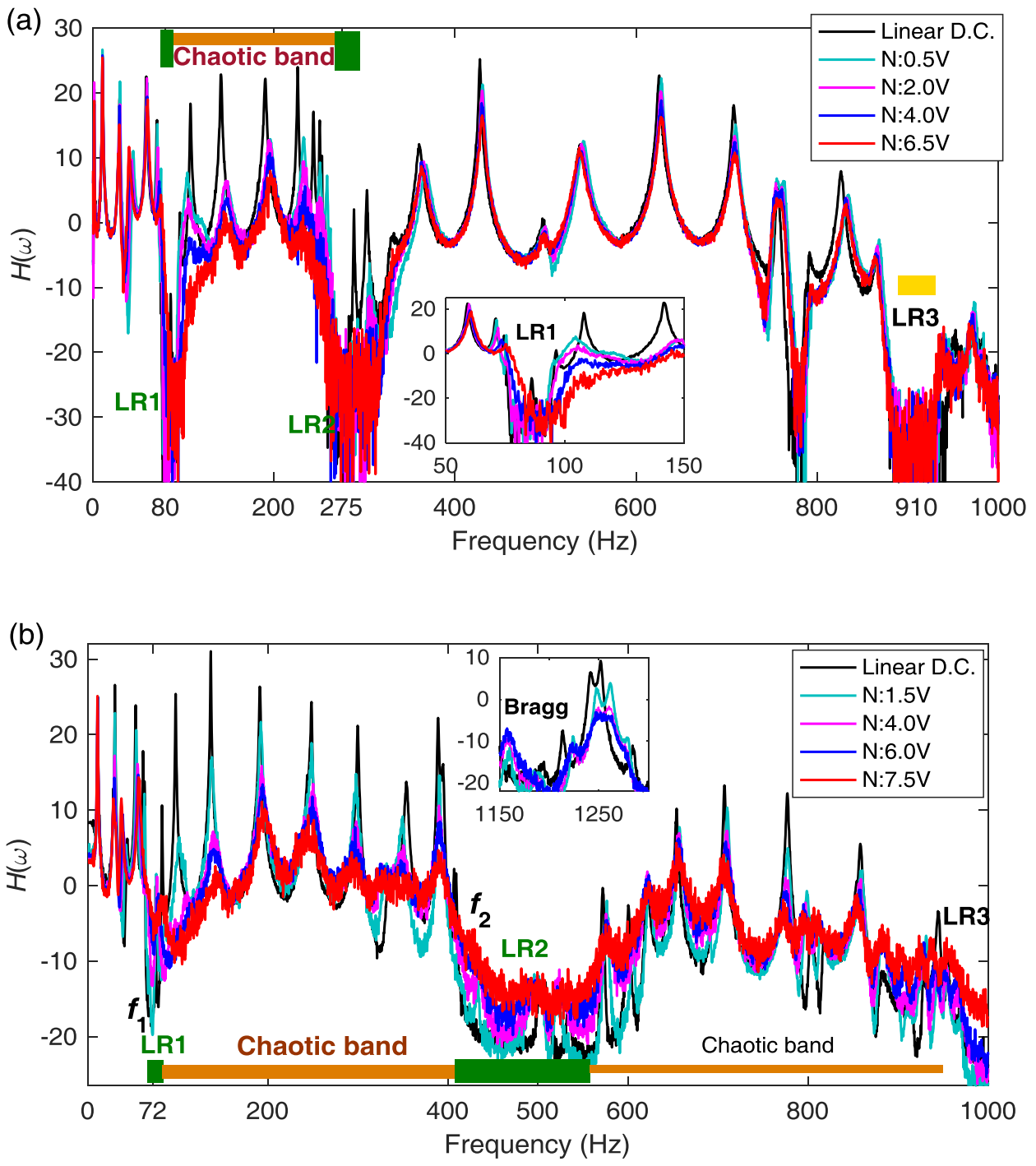

FIG. 8 Transfer function $H(\omega)$ under different excitation levels. (a) NAM1. (b) NAM2. The shaded frequency ranges represent bandgaps and chaotic bands. Using the average amplitude in 50-100 Hz (near LR1) to evaluate different excitation levels. NAM1: {0.5 V, 0.7446 μm}, {2.0 V, 2.7458 μm}, {4.0 V, 5.4454 μm}, {6.5 V, 8.7768 μm}; NAM2: {1.5 V, 0.8734 μm}, {4.0 V, 2.372 μm}, {6.0 V, 3.502 μm}, {7.5 V, 11.7 μm}. Linear D.C. represents the decoupled case; it is the linear state of the metamaterial. N-$x$V denotes the voltage of the amplifier used to drive the NAM.

As shown in FIG. 8, for LAMs, the transmission losses induced by the bandgaps reach 30 dB. However, the passbands consist of dense resonances that amplify waves more than 20 dB. As characterized by $H(\omega)$ of NAM1 (see FIG. 8a), increasing the driving amplitude shifts LR1 to higher frequencies, which is accordant with the dispersion results. Moreover, wave transmissions of the resonances in the second passbands (LR1-LR2) decrease by 15-30 dB when strong nonlinearity appears (6.5V). As clarified [46], such significant wave attenuation is induced by the chaotic band effect because the second passband become chaotic band. In addition, in the third passband, only resonances near LR2 are suppressed, and the maximum $H(\omega)$ of NAM1 under 6.5V and LAM1 reach 16 dB and 25 dB, respectively. The wave suppressions/attenuations of NAM1 under strongly nonlinear regime mainly appear in 70-300 Hz that consists of chaotic band and bandgaps; its generalized width $\gamma$>3.3. Those features of NAM1 are consistent with the short distance $\Delta\omega$ case of the triatomic model (see FIG. 2a,c), which verifies that the wave attenuations in the chaotic band are controlled by the bridging coupling pair NLR1-NLR2. The energy transfer between oscillator-1 and oscillator-2 shares their negative mass properties in the chaotic band to suppress the nonlinear resonances. In theory, there are bridging couplings between NLR3 with other bandgaps. However, much stronger nonlinearity is required to generate obvious interactions because of the long frequency distances.

When the distance between NLR1 and NLR2 increases, as predicted by the remote bridging couplings of the triatomic model: the chaotic band suppressing nonlinear resonances become broader with the second passband; the maximum $H(\omega)$ of LAM in the third passband decreases; and the third passband of NAM can also suppress the resonances. As shown in FIG. 8b, $H(\omega)$ of NAM2 features same properties as predictions of the bridging coupling: in the second passband (the chaotic band), the resonant transmissions under the strongly nonlinear state (7.5V) decrease by 15-35 dB relative to the decoupled linear state; NLR1 is closed and shifted upwards; the maximum $H(\omega)$ of NAM2 under 6.0 V and LAM2 are just 6 dB and 13 dB, which is much lower than that of NAM1. Therefore, NLR1 and NLR2 of NAM2 are bridging coupled. Besides, different from the properties of the third passband of NAM1, resonances far from NLR2 and near NLR3 in NAM2 are also suppressed. This effect even spreads to the resonances (near 1250 Hz) above the Bragg bandgap (see the small panel in FIG. 8b). Parts of this suppression arises from the farther bridging pair NLR1-NLR2 under strongly nonlinear state, but the bridging coupling between NLR2 and NLR3 also acts on it. Therefore, bridging couplings between multiple NLR bandgaps are desirable. Even if neglecting the Bragg bandgap and higher frequencies, those bridging couplings in NAM2 bring wave suppressions/attenuations in an ultra-low frequency and ultra-broad band 60-1000 Hz. The generalized bandwidth reaches $\gamma$ >15.6 that is about five times broader than NAM1. Experimental phenomena demonstrate that increasing the distance between bridging NLR bandgaps can manipulate the vibration suppression bandwidth and efficiency in their common chaotic band.

## C. Bifurcations of periodic solutions

To confirm the responses at the nonlinear resonances are chaotic, we find periodic solutions of NAM2's FE model by combining the harmonic average method and the perturbation continuation approach. Stabilities of the periodic solutions are determined by the eigenvalues of the Jacobian matrix (see Appendix). Driving force $F$ is applied at point E. Dissipations of the pyramid rubbers and the structural damping are taken into consideration (see Appendix).

We find periodic solutions in three typical ranges in the second passband: the lower (near LR1, 80-170 Hz), the middle (210-245 Hz) and the upper (near LR2, 350-410 Hz) parts, as shown in FIG. 9. Under $F$ =1 N, the nonlinear resonances characterize three branches bending rightwards, but all the resonant peaks are unstable (or the stable and unstable ranges appears alternately) and their amplitudes are much smaller than the linear resonant peaks. On this occasion, the root branches of the nonlinear resonances are stable and coincide with linear results, which indicate quasi-periodic solutions. These phenomena like the studies made on diatomic/tetratomic models in Ref. [45]. Moreover, at the lower and the upper parts near the bandgaps, all solutions become unstable indicating that nonlinear effects present in both oscillator-1 and oscillator-2.

Further increasing the driving force to $F$ =5 N (FIG. 9b), not only the resonant peaks but also their roots become unstable; the unstable ranges enlarge and solutions of the middle part in this passband become unstable. For example, in 80-170 Hz under $F$ =5 N, stable solutions can only be found in some narrow frequency intervals; other ranges are consist of complex bifurcation branches formed by unstable solutions. These frequency ranges consisting of unstable periodic solutions or featuring the alternate stable and unstable solutions lead to chaotic responses [45]. Therefore, the whole second passband becomes chaotic band under strongly nonlinear state of NAMs. Furthermore, increasing the driving force reduces the generalized amplitude $s/F$ significantly. Accordingly, stronger nonlinearity results in greater transmission losses in the chaotic band.

These analyses of periodic solutions confirm that the bridging coupling pair NLR1-NLR2 makes the passband between them a chaotic band suppressing wave transmissions.

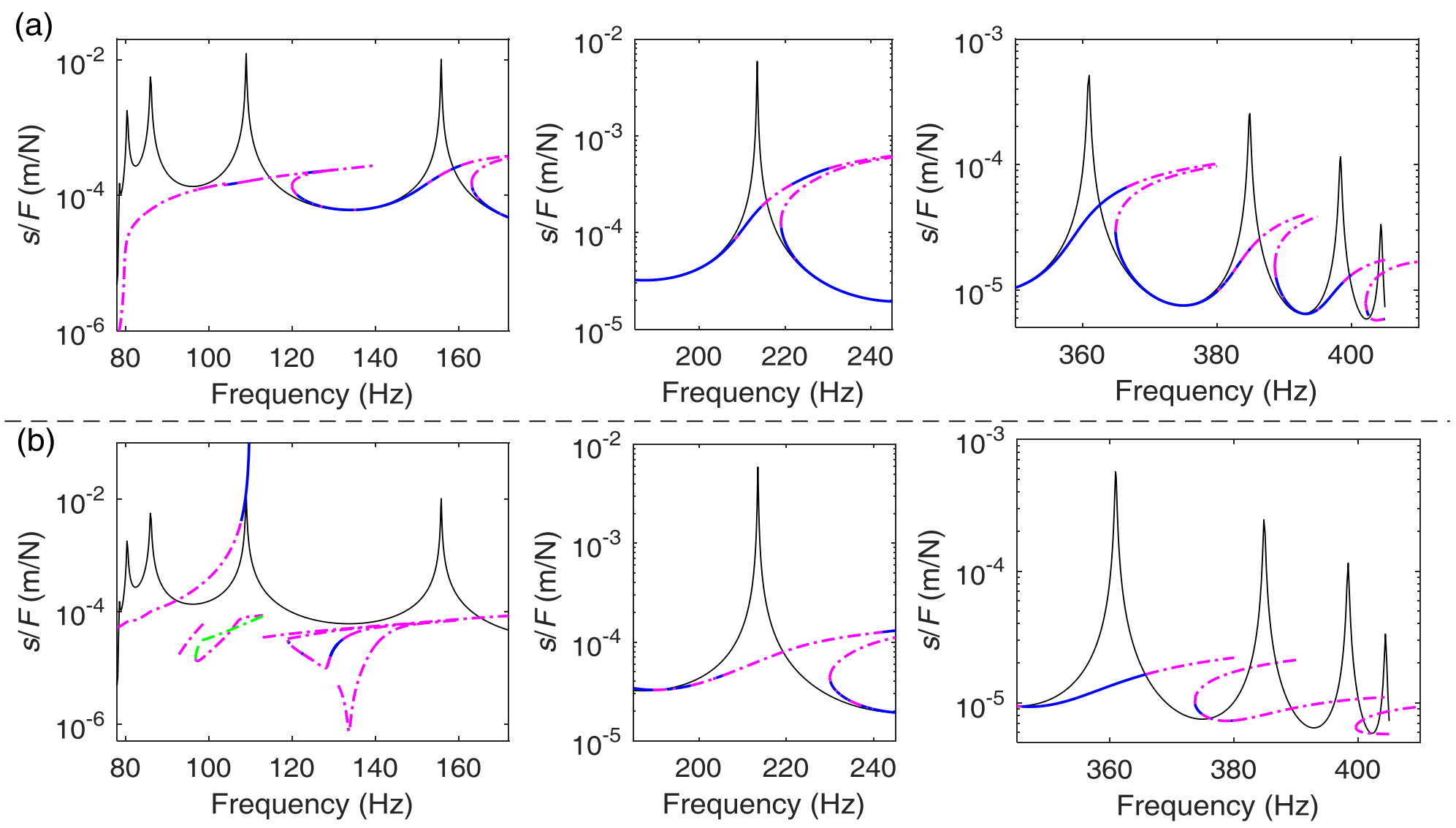

FIG. 9 Bifurcations of periodic solutions found by the perturbation continuation approach. (a) $F$ = 1 N; (b) $F$ = 5 N. All the solutions are convergent. The solid black curves are solutions of LAM2; the solid blue (dashed magenta/green) curves represent stable (unstable) periodic solutions of NAM2.

# IV. CONCLUSIONS

Chaotic band in nonlinear acoustic metamaterial enables the ultra-low frequency, ultra-broad-band and highly efficient wave reductions (the double-ultra effect) [46]. This work reports the remote bridging coupling of nonlinear locally resonant (NLR) bandgaps for efficient manipulations of chaotic bands.

The finite triatomic NAM model whose cell contains two nonlinearly coupled resonators is created to study the couplings of NLR bandgaps. This coupling features broadband sharing of the negative mass through energy transfer between resonators. We find numerically that the passband between two NLR bandgaps becomes a chaotic band, and increasing the frequency distance between the two NLR bandgaps can enhancing the efficiency meanwhile expanding the bandwidth for the wave suppression in the chaotic band. This remote interaction is essentially different from the adjacent bandgap couplings, so it is highlighted as Bridging Coupling.

To demonstrate the bridging coupling of NLR bandgaps, two NAM beams with different distances of NLR bandgaps are fabricated. We experimentally verify that the transmission losses and the bandwidth of the chaotic band (between and near NLR bandgaps) can be modulated by the distance of NLR bandgaps. Ultra-low frequency and ultra-broad-band wave suppressions/ attenuations are achieved again (the generalized bandwidth $\gamma$ >15.6). With the nonlinear FE model of the NAM beam, periodic solutions found by continuation approach confirm the broadband responses become chaotic waves whose transmissions are amplitude-dependent. Therefore, bridging couplings of NLR bandgaps are demonstrated to be an efficient mechanism to manipulate the chaotic band under the limited nonlinear strength in practice.

The double-ultra wave suppressions enabled by chaotic bands in NAMs are desired in widely applications, such as vibration and noise control, broadband metadevices. This work reports and demonstrates the remote bridging coupling of nonlinear resonant bandgaps for generating and manipulating the chaotic band conveniently and efficiently. Our study extends the content of NAMs and more nonlinear effects are anticipated based on this mechanism.

# ACKNOWLEDGEMENTS

This research was funded by the National Natural Science Foundation of China (Projects Number 51405502, number 51275519).

# APPENDIX: NONLINEAR FINITE ELEMENT MODEL

Modal superposition method is adopted to model the flexural vibration of the oscillators. The first two modes of oscillator-1 and the first mode of oscillator-2 are considered. Transverse displacements $w_1(x)$ and $w_2(x)$ of Oscillator-1 and Oscillator-2 are respectively expressed by

$$\begin{aligned} w_1(x) &= W_{11}(x)q_{11}(t) + W_{12}(x)q_{12}(t) \\ w_2(x) &= W_{21}(x)q_{21}(t) \end{aligned} \tag{10}$$

Herein, $x$ denotes the distance to the fixed point; $q_{ij}$ is the modal coordinate; $W_{ij}(x)$ represents the modal displacement given by

$$\begin{aligned} W_{ij}(x) &= \sin\beta_{ij}x - \sinh\beta_{ij}x \\ &+ \frac{\cos\beta_{ij}L_i + \cosh\beta_{ij}L_i}{\sin\beta_{ij}L_i - \sinh\beta_{ij}L_i}(\cos\beta_{ij}x - \cosh\beta_{ij}x) \end{aligned} \tag{11}$$

where, $\cos\beta_{ij}L_i \cosh\beta_{ij}L_i = -1$. Therefore, $\beta_{i1}L_i = 1.875$, $\beta_{i2}L_i = 4.694$. The kinetic energy $T_r$ and the strain energy $U_r$ of oscillators in a cell are given by

$$\begin{aligned} T_r &= \frac{1}{2}m_0\dot{w}_0^2 + \frac{1}{2}J_0\dot{\theta}_0^2 + \frac{1}{2}\int_0^{L_1}\rho_1 A_1(\dot{w}_0 + \dot{w}_1 + x\dot{\theta}_0)^2 dx \\ &+ \frac{1}{2}m_{t1}[\dot{w}_0 + \dot{w}_1(L_1) + L_1\dot{\theta}_0]^2 + \frac{1}{2}J_{t1}[\dot{\theta}_0 + \dot{\theta}_1(L_1)]^2 \\ &+ \frac{1}{2}\int_0^{L_2}\rho_2 A_2(\dot{w}_0 + \dot{w}_2 + x\dot{\theta}_0)^2 dx + \\ &\frac{1}{2}J_{t2}[\dot{\theta}_0 + \dot{\theta}_2(L_2)]^2 + \frac{1}{2}m_{t2}[\dot{w}_0 + \dot{w}_2(L_2) + L_2\dot{\theta}_0]^2 \end{aligned} \tag{12}$$

$$\begin{aligned} U_r &= \frac{1}{2}\int_0^{L_1}E_1 I_1(w_{1,xx})^2 dx + \frac{1}{2}\int_0^{L_2}E_2 I_2(w_{2,xx})^2 dx \\ &+ \frac{1}{2}k_1\Delta^2 + \frac{1}{4}k_n\Delta^4 \\ \Delta &= w_1(L_1,t) - w_2(L_2,t), \quad \theta_i = \partial w_i/\partial x \end{aligned} \tag{13}$$

where $(\cdot)_x=\partial/\partial x$ and $\dot{X}=\partial X/\partial t$; $w_0$ and $\theta_0$ are deflection and angle displacements of the primary beam at the fixed point respectively. All parameters are listed in Table 1. By substituting Equ. (11) into Equs. (12) and (13) we obtain

$$T_r = \frac{1}{2}\dot{\tilde{\mathbf{q}}}_r \tilde{\mathbf{m}}_r \dot{\tilde{\mathbf{q}}}_r^{\mathrm{T}}, \quad U_r = \frac{1}{2}\tilde{\mathbf{q}}_r \tilde{\mathbf{k}}_{rL} \tilde{\mathbf{q}}_r^{\mathrm{T}} + \frac{1}{4}k_n \Delta^4 \tag{14}$$

here the vector $\tilde{\mathbf{q}}_r = [w_0 \quad \theta_0 \quad q_{11} \quad q_{12} \quad q_{21}]^{\mathrm{T}}$; $\tilde{\mathbf{m}}_r$ and $\tilde{\mathbf{k}}_{rL}$ are linear mass and stiffness matrices, respectively. The derivation yields,

$$\tilde{\mathbf{m}}_r = \begin{bmatrix} m_{r,11} & m_{r,12} & m_{r,13} & m_{r,14} & m_{r,15} \\ & m_{r,22} & m_{r,23} & m_{r,24} & m_{r,25} \\ & & m_{r,33} & m_{r,34} & 0 \\ & \text{sym} & & m_{r,44} & 0 \\ & & & & m_{r,55} \end{bmatrix},$$
$$\tilde{\mathbf{k}}_{rL} = \begin{bmatrix} 0 & 0 & 0 & 0 & 0 \\ & 0 & 0 & 0 & 0 \\ & & k_{r,33} & k_{r,34} & k_{r,35} \\ & \text{sym} & & k_{r,44} & k_{r,45} \\ & & & & k_{r,55} \end{bmatrix} \tag{15}$$

Elements in $\tilde{\mathbf{m}}_r$ and $\tilde{\mathbf{k}}_{rL}$ are not repeated here. The nonlinear term in Equ. (14) can be simplified by using the formula

$$\begin{aligned} \Delta = W_{11}(L_1)q_{11} + W_{12}(L_1)q_{12} - W_{21}(L_2)q_{21} \\ + (L_1 - L_2)\theta_0 = q_d \end{aligned} \tag{16}$$

A transformation yields

$$\begin{aligned} q_{21} = [(L_1 - L_2)\theta_0 + W_{11}(L_1)q_{11} \\ + W_{12}(L_1)q_{12} - q_d]/W_{21}(L_2) \end{aligned} \tag{17}$$

Defining a new vector $\mathbf{q}_r = [w_0 \quad \theta_0 \quad q_{11} \quad q_{12} \quad q_d]^{\mathrm{T}}$, we obtain $\tilde{\mathbf{q}}_r = \mathbf{A}_r \mathbf{q}_r$ and

$$\mathbf{A}_r = \begin{bmatrix} 1 & & & & \\ 0 & 1 & & & \\ 0 & 0 & 1 & 0 & 0 \\ 0 & 0 & 0 & 1 & 0 \\ 0 & \frac{L_1 - L_2}{W_{21}(L_2)} & \frac{W_{11}(L_1)}{W_{21}(L_2)} & \frac{W_{12}(L_1)}{W_{21}(L_2)} & \frac{-1}{W_{21}(L_2)} \end{bmatrix} \tag{18}$$

With the vector transformation, matrices for the mass, the linear stiffness and the nonlinear stiffness coefficient are

$$\begin{aligned} &\mathbf{m}_r = \mathbf{A}_r^{\mathrm{T}} \tilde{\mathbf{m}}_r \mathbf{A}_r, \\ &\mathbf{k}_{rL} = \mathbf{A}_r^{\mathrm{T}} \tilde{\mathbf{k}}_{rL} \mathbf{A}_r, \\ &\mathbf{k}_{rN} = \mathbf{0} \quad \text{and} \quad k_{rN,55} = k_n \end{aligned} \tag{19}$$

If the dissipation arising from the pyramid rubber is approximated as the linear viscous damping, the virtual work made by a rubber is given by

$$\delta W_d = -c\dot{\Delta}\delta\Delta = -c\dot{q}_d \delta q_d \tag{20}$$

Therefore, the motion equation of the oscillators in a cell is expressed as

$$\mathbf{m}_r \ddot{\mathbf{q}}_r + \mathbf{c}_r \dot{\mathbf{q}}_r + \mathbf{k}_{rL} \mathbf{q}_r + \mathbf{k}_{rN} \mathbf{q}_r^3 = \mathbf{f} \tag{21}$$

where $\mathbf{c}_r$ is the damping matrix of the oscillators, and only the element $c_{r,55}=c$ is nonzero.

As verified, the geometrical nonlinearity of the primary beam is neglectable [46]. Furthermore, we establish the finite element (FE) model of the whole NAM beam by following a standard FE procedure. The motion equation of the whole NAM beam is given by

$$\mathbf{M}\ddot{\mathbf{q}} + \mathbf{K}\mathbf{q} + \mathbf{C}\dot{\mathbf{q}} + \mathbf{N}\mathbf{q}^3 = \mathbf{F}\sin\omega t \tag{22}$$

Where $\mathbf{q}$ and $\mathbf{F}$ are vectors for all nodes' displacement and force, respectively; $\omega$ denotes the driving frequency at point E; $\mathbf{M}$, $\mathbf{K}$, $\mathbf{C}$, $\mathbf{N}$ are system's mass, linear stiffness, damping and nonlinear stiffness coefficient matrices, respectively. Besides the damping of the pyramid rubber, the weak dissipation arising from the deformation of the structure is also considered, so the matrix $\mathbf{C}$ is superposed by two damping sources,

$$\mathbf{C} = \mathbf{C}_r + \eta_s \mathbf{K}/\omega \tag{23}$$

In Equ. (23), $\mathbf{C}_r$ is expanded with the damping matrix $\mathbf{c}_r$ in Equ. (21); $\eta_s$ symbolizes the damping coefficient of the structure. In theoretical analyses, we specify $c$=0.01 and $\eta_s=1\times10^{-5}$.

Harmonic average approach is adopted to solve the approximate frequency responses (the periodic solutions) of Equ. (22) and determine the stabilities of the solutions. The steps of this algorithm are detailed in Ref. [45]. By defining,

$$\mathbf{q} = \mathbf{u}\cos\omega t + \mathbf{v}\sin\omega t \tag{24}$$

The periodic solution is determined by the algebraic equations below

$$\begin{cases} (\mathbf{K} - \omega^2\mathbf{M})\mathbf{v} - \omega\mathbf{C}\mathbf{u} + 3\mathbf{N}\left((\mathbf{v}^2 + \mathbf{u}^2)\mathbf{v}\right)/4 = 0 \\ (\mathbf{K} - \omega^2\mathbf{M})\mathbf{u} + \omega\mathbf{C}\mathbf{v} + 3\mathbf{N}\left((\mathbf{v}^2 + \mathbf{u}^2)\mathbf{u}\right)/4 = \mathbf{F} \end{cases} \tag{25}$$

The Jacobian matrix $\mathbf{J}$ used to analyze the stabilities of the solutions is given by

$$\mathbf{J}=\begin{bmatrix}\mathbf{M}^{-1} & \mathbf{0}\\ \mathbf{0} & -\mathbf{M}^{-1}\end{bmatrix}\begin{bmatrix}-\omega\mathbf{C}+3\mathbf{N}\circ[[\mathbf{uv}]]/2 & \mathbf{K}-\omega^2\mathbf{M}+3\mathbf{N}\circ[[3\mathbf{v}^2+\mathbf{u}^2]]/4\\ \mathbf{K}-\omega^2\mathbf{M}+3\mathbf{N}\circ[[3\mathbf{u}^2+\mathbf{v}^2]]/4 & \omega\mathbf{C}+3\mathbf{N}\circ[[\mathbf{uv}]]/2\end{bmatrix} \tag{26}$$

Here $[[\mathbf{x}]]=[\mathbf{x}\quad \mathbf{x}\quad \cdots\quad \mathbf{x}]^{\mathrm{T}}$ is the algebraic evaluation for expanding a vector **x** into a square matrix [[**x**]]. $\mathbf{A}\circ\mathbf{B}=[a_{ij}b_{ij}]$ is defined as the element product of two matrices **A** and **B**.

Moreover, perturbation continuation algorithm is adopted to find convergent solutions of Equ. (25). Substituting a solution into Equ. (26) leads to the matrix **J.**

If there are eigenvalues of **J** having positive real parts, the corresponding solution is unstable, otherwise it is stable. Due to the inevitable numerical errors, we use a positive value $\mu_c=1\times10^{-8}$ instead of $\mu_c=0$ as the critical value for the real parts of eigenvalues to determine the stability.